\documentclass[oneside]{article}
\usepackage[T1]{fontenc}
\usepackage{authblk}
\usepackage{float}
\usepackage{amsmath}
\usepackage{graphicx}
\usepackage{xfrac}
\usepackage[english]{babel}
\usepackage{lmodern}
\usepackage[T1]{fontenc}
\usepackage[latin9]{inputenc}
\usepackage{mathtools}
\usepackage{graphicx}
\usepackage{esint}
\usepackage[unicode=true,
 bookmarks=false,
breaklinks=false,pdfborder={0 0 1},backref=false,colorlinks=false]
 {hyperref}
\usepackage[a4paper,margin=2.0cm]{geometry}

\usepackage{datetime}
\newdate{date}{28}{01}{2015}

\title{Non-linear Langevin model for the early-stage dynamics of electrospinning
jets.}

\author[1]{Marco Lauricella}
\author[1]{Giuseppe Pontrelli}
\author[2,3]{Dario Pisignano}
\author[1]{Sauro Succi \thanks{Electronic address: \texttt{succi@iac.cnr.it}; Corresponding author}}

\affil[1]{Istituto per le Applicazioni del Calcolo CNR, Via dei Taurini 19, 00185 Rome, Italy}
\affil[2]{Dipartimento di Matematica e Fisica Ennio De Giorgi,University of Salento, via Arnesano, 73100 Lecce, Italy}
\affil[3]{Istituto Nanoscienze CNR, Via Arnesano 16, 73100 Lecce, Italy}

\date{\displaydate{date}}

\begin{document}

\maketitle

\begin{abstract}
We present a non-linear Langevin model to investigate the early-stage
dynamics of electrified polymer jets in electrospinning experiments. 
In particular, we study the effects of air drag force on the uniaxial 
elongation of the charged jet, right after ejection from the nozzle. 
Numerical simulations show that the elongation of the jet filament 
close to the injection point is significantly affected by the 
non-linear drag exerted by the surrounding air.
These result provide useful insights for the optimal design of 
current and future electrospinning experiments. 
\end{abstract}

\section{Introduction}

The dynamics of charged fluids under the effect of an external electrostatic 
fields represents a major theme of non-equilibrium thermodynamics 
and statistical mechanics
\cite{baus1980statistical,HansenChargedFluids,HansenNonequilibrium,carof2014coarse}.

Charged liquid jets may develop several types of instabilities,
depending on the relative strength of the various forces acting upon
them, primarily electrostatic Coulomb self-repulsion, viscoelastic
drag, surface tension effects and dissipative forces due to the
interaction with the surrounding environment. 
Since these instabilities are central to many industrial processes, including
the production of ultrathin fibers via the so-called electrospinning, they
have been analyzed since the late 1960s \cite{taylor1969electrically}.
However, due to inherent difficulties associated with the underlying long-range
many-body problem, a thorough theoretical understanding is still lacking,
whence an important role for computer simulation.

In the recent years, the production of ultrathin nanofibers has found
increasing applications in micro and nanoengineering and life sciences
as well \cite{pisignanoelectrospinning,yarin2014fundamentals}.
In the electrospinning process, electrospun nanofibers are produced
at laboratory scale by the uniaxial stretching of a jet, which is
ejected at the nozzle from the surface of a charged polymer solution.
This initial elongation of a jet is produced by applying an externally
electrostatic field. An intense electric field 
(typically $10^{5}-10^{6}\text{V}\cdot\text{m}^{-1}$)
is applied between the spinneret and an oppositely charged 
collector, usually placed at about $20$ cm from the injector. 
Electrospinning involves mainly two sequential stages in the uniaxial elongation
of the extruded polymer jet: an initial growth stage, in which
the electric field stretches the jet along a straight path away from
the nozzle of the ejecting apparatus, and a second stage characterized
by a bending instability induced by small perturbations, which misalign
the jet away from its axis of elongation. These small disturbances may
originate from Coulomb repulsion on different portions of the
jet, as well as from mechanical vibrations at the nozzle or aerodynamic
perturbations within the experimental apparatus. 
As a consequence,
the jet path length between the nozzle and the collector increases,
and the stream cross-section undergoes a further decrease. The prime
goal of electrospinning experiments is to minimize the radius of the
collected fibers. By a simple argument of mass conservation, this
is tantamount to maximizing the jet length by the time it reaches
the collecting plane. Consequently, the bending instability is a desirable effect, as 
it produces a higher surface-area-to-volume ratio of the jet, which is 
transferred to the resulting nanofibers. 
By the same argument, it is therefore of interest
to minimize the length of the initial stable jet region. 

Simulation models provide a useful tool to elucidate the phenomenon
and provide valuable information for the design of future electrospinning
experiments. Numerical simulations enhance the capability of predicting
the key process parameters and exert a better control on the resulting
nanofiber structure. In recent years, with a renewed interest in nanotechnology,
electrospinning studies attracted the attention of many researchers
both from modeling and experimental points of view \cite{reneker2000bending,persano2013industrial}.
These models treat the jet filament either as a charged continuum
fluid, or as a series of discrete elements (\textit{beads}) obeying the equations
of Newtonian mechanics\cite{yarin2014fundamentals}.
The latter standpoint is the one taken in this work.
Each bead is subject to different types of interactions, namely long-range
Coulomb repulsion, viscoelastic drag and the external electric field.
The main aim of such models is to investigate the complexity of the
resulting dynamics and provide the optimal set of parameters driving
the process. Actually, due to the large number of experimental parameters,
electrified jets are still treated via empirical approaches. The effect
of fast-oscillating loads on the bending instability, have been explored
in a modeling and computational study \cite{coluzza2014ultrathin}.
On the other hand, the effect of air drag in electrospinning process
has been addressed only very recently \cite{lauricella2014electrospinning},
even though there is experimental evidence that the air drag affects
the dynamics of the nanofiber via a non-linear dependence on the jet
geometry\cite{spinning1991science}.

Here, we investigate the uniaxial elongation of an electrified polymer
jet in the early-stage of its dynamics at the nozzle of the ejecting
apparatus and under stochastic-dissipative force. 
This perturbation effect is modeled by a Langevin approach. 
In particular,
we assume that a Brownian term can adequately reproduce a stationary
perturbation related to multiple simultaneous tiny impacts along the
direction of the jet elongation axis, as in the case of an air drag
force generated by the motion of a polymer jet through a gaseous medium.
Relations between air drag force and Brownian motion were proposed
in literature starting from experimental observations\cite{antonia1980measurements,sinha2010meltblowing}.
In a recent work, we extended the one-dimensional bead-spring model
developed by Pontrelli \textit{et al.}\cite{pontrelli2014electrospinning},
to include a linear dissipative-perturbing Langevin force which models
the effects of the air drag force. This is accomplished by adding
a random and a dissipative force to the equations of motion, and further
assuming that the two terms obey the fluctuation-dissipation theorem
\cite{gillespie2012simple}. As a result, the system was described
by a Langevin-like linear stochastic differential equation \cite{lauricella2014electrospinning}.

Based on empirical evidences, in this paper we extend the previous
approach to include non-linear effects due to air disturbances (see
Sec 2). The resulting non-linear Langevin equation is numerically
integrated to investigate different dynamical regimes under the effect
of systematic and random air perturbations (see Sec 3).

\section{The mathematical model}

\begin{figure}[H]
\begin{centering}
\includegraphics[scale=0.50]{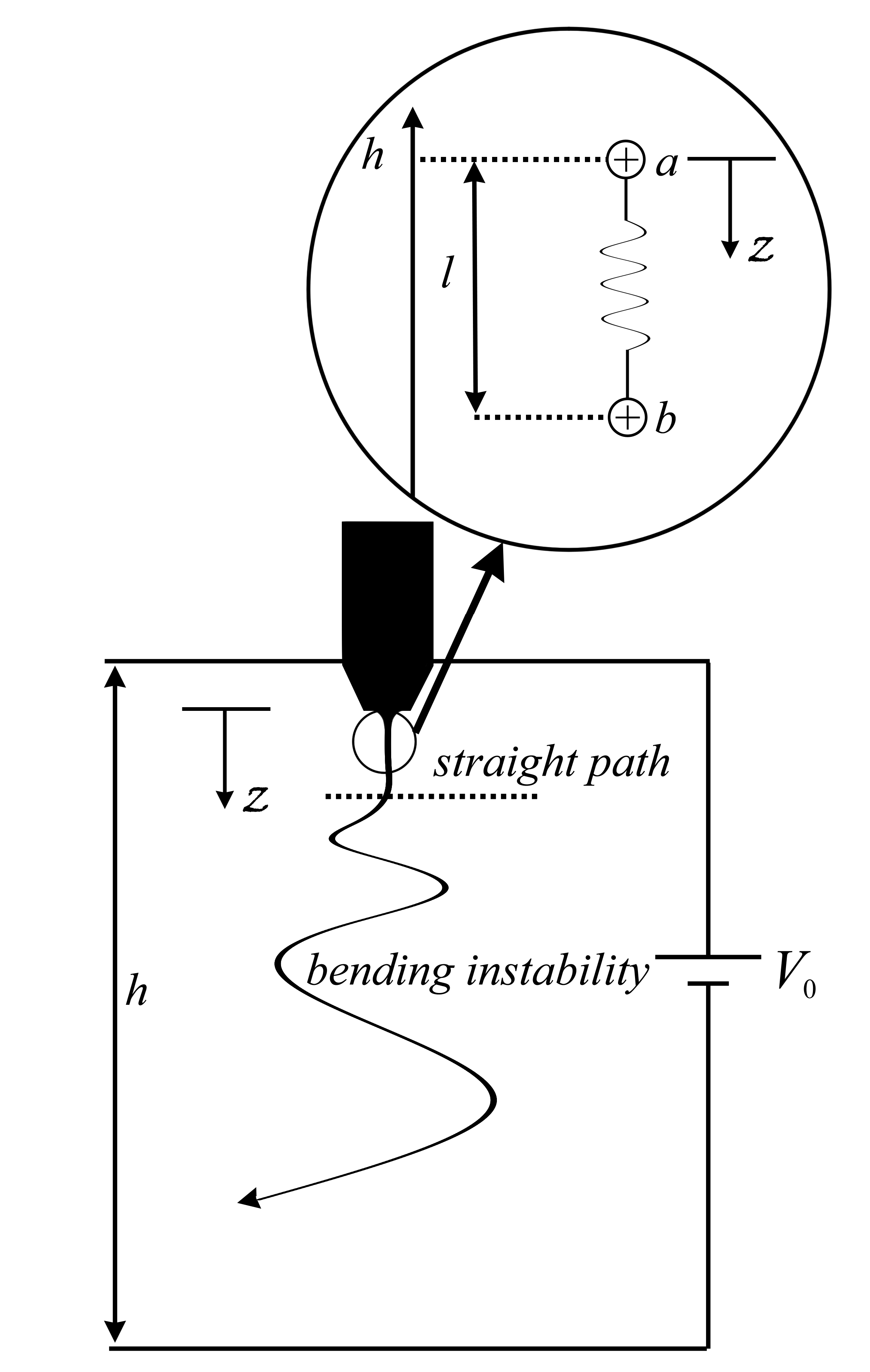} 
\par\end{centering}

\protect\caption{{\small{}Schematic drawing of the electrospinning process (not in
scale), showing $h$ the distance between the collector plate and
the injection point (nozzle), $V_{0}$ the applied voltage between
these two elements, and the $z$ reference axis whose origin is fixed
at the injection point.}}

\label{Fig:schema-esperimento} 
\end{figure}

Let us consider a rectilinear electrified viscoelastic jet in a typical
electrospinning experiment. In order to model the stretching, we represent
the filament by a viscoelastic dumbbell (or dimer) with two charged
beads ($a,b$) with the same mass $m$ and charge $e$ (not to be confused
with the electron charge) located a distance $l$ apart (Fig. \ref{Fig:schema-esperimento}). 
One of the two beads ($a$) is held fixed to the nozzle, while the other
one ($b$) is free to move under the effect of the various forces.
The dumbbell $ab$ represents a schematic model for the jet. We denote
by $h$ the distance of the collector plate from the injection point
and by $V_{0}$ the applied voltage between them. The bead $b$ is
subject to three different forces (gravity and surface tension are
neglected): the external electrical field $V_{0}/h$, the Coulomb
repulsive force between the two beads, and the viscoelastic force.
As pointed out elsewhere \cite{lauricella2014electrospinning,pontrelli2014electrospinning,reneker2000bending},
the competition between Coulomb and viscoelastic forces characterizes
the early-stage of the jet elongation, while the second stage is dominated
by the external electrical force, driving the fluid filament towards
the collector.

The combined action of these three forces governs the elongation of
the jet according to the following equation{\small{}\cite{reneker2000bending}}:

\begin{equation}
m\frac{d\upsilon}{dt}=\frac{e^{2}}{l^{2}}+\frac{eV_{0}}{h}-\pi r^{2}\sigma,\label{eq:vel-ode}
\end{equation}
where $\upsilon$ is the velocity of the bead $b$, $t$ the time,
$r$ the cross-sectional radius of the filament, $\pi r^{2}\sigma$
the force pulling the bead $b$ back to $a$, with $\sigma$ the stress
of the viscoelastic force. Assuming a viscoelastic Maxwellian liquid 
\cite{yarin2014fundamentals} the time evolution of the stress $\sigma$
related to the viscoelastic force is provided by the equation:

\begin{equation}
\frac{d\sigma}{dt}=G\frac{dl}{ldt}-\frac{G}{\mu}\sigma,\label{eq:stress-ode}
\end{equation}
where $G$ is the elastic modulus, and $\mu$ the viscosity of the
fluid. 
The velocity $\upsilon$ satisfies the kinematic relation:

\begin{equation}
\frac{dl}{dt}=\upsilon.\label{eq:pos-ode}
\end{equation}

To include air drag effects on the dynamics of electrified jets, we
add a random and dissipative term into Eq (\ref{eq:vel-ode}).
Denoted $D_{\upsilon}$ the diffusion coefficient in velocity
space and $\alpha$ the friction parameter, we assume that the dissipative
term has the form $\alpha W$, while the random force term the form
$\sqrt{2D_{\upsilon}}\eta\left(t\right)$, where $\eta\left(t\right)$
is a nowhere differentiable stochastic process with $<\eta\left(t_{1}\right)\eta\left(t_{2}\right)>=\delta\left(\left|t_{2}-t_{1}\right|\right)$,
and $<\eta\left(t\right)>=0$. As a first approximation, we have taken
$\alpha$ constant \cite{lauricella2014electrospinning}. Adding
these two force terms in Eq (\ref{eq:vel-ode}), we obtain a Langevin-like
stochastic differential equation:

\begin{equation}
m\frac{d\upsilon}{dt}=\frac{e^{2}}{l^{2}}+\frac{eV_{0}}{h}-\pi r^{2}\sigma-m\alpha\upsilon+\sqrt{2m^{2}D_{\upsilon}}\eta\left(t\right)\label{eq:SDE-dimen}
\end{equation}

A dependence of the dissipative term both on the velocity and
on the length of nanofibers was experimentally reported 
\cite{spinning1991science,sinha2010meltblowing,yarin2014fundamentals},
and the following empirical formula for the dissipative air drag force
was proposed:

\begin{equation}
f_{air}=-m\alpha l^{0.905}\upsilon^{1.19},\label{eq:emp-air}
\end{equation}

where $\alpha$ denotes an empirical factor
depending on the air density and kinematic viscosity of gaseous medium.

Therefore, we propose here a more comprehensive model for the Langevin-like
stochastic differential equation (\ref{eq:ODE}.3) which takes the
form:

\begin{equation}
m\frac{d\upsilon}{dt}=\frac{e^{2}}{l^{2}}+\frac{eV_{0}}{h}-\pi r^{2}\sigma-m\alpha l^{\, S}\upsilon^{1+P}+\sqrt{2m^{2}D_{\upsilon}}\eta\left(t\right),\label{eq:SDE-dimen-new}
\end{equation}
where the parameter $P$ denotes the super-linearity of the dissipation
term, and $S$ accounts for a sub-linear dependence of the dissipative
term on the jet length. It is worth observing that Eq (\ref{eq:SDE-dimen-new})
reduces to Eq (\ref{eq:SDE-dimen}) in the limit $P,S\rightarrow0$.

All the variables and equations are recast in a more convenient non-dimensional
form \cite{reneker2000bending}. To this aim, we define a length
scale $L=\left(e^{2}/\pi r_{0}^{2}G\right)^{1/2}$, at which Coulomb
repulsion matches the reference viscoelastic stress $G$, 
being $r_{0}$ the initial radius, and $\tau=\mu/G$ a relaxation time.
By defining

\begin{equation}
\begin{alignedat}{1}\bar{t} & =\frac{t}{\tau}\\
\bar{l} & =\frac{l}{L}\\
\bar{W} & =\upsilon\cdot\frac{\tau}{L}\\
\bar{\alpha} & =\alpha\cdot\tau\, L^{S}\left(\frac{L}{\tau}\right)^{P}\\
\bar{D}_{\upsilon} & =D_{\upsilon}\cdot\frac{\tau^{3}}{L^{2}}
\end{alignedat}
\end{equation}
and applying the conservation of the jet volume, $\pi r^{2}\, l=\pi r_{0}^{2}L$,
the equations of motion in terms of non-dimensional (barred) variables
take the following form: 
\begin{subequations} 
\begin{align}
\frac{d\bar{l}}{d\bar{t}} & =\bar{W}\\
\frac{d\bar{\sigma}}{d\bar{t}} & =\frac{\bar{W}}{\bar{l}}-\bar{\sigma}\\
\frac{d\bar{W}}{d\bar{t}} & =\frac{Q}{\bar{l}^{2}}+V-F_{ve}\frac{\bar{\sigma}}{\bar{l}}-\bar{\alpha}\bar{l}^{\, S}\bar{W}^{1+P}+\sqrt{2\bar{D}_{\upsilon}}\eta\left(\bar{t}\right)\label{eq:SDE}
\end{align}
\label{eq:ODE} \end{subequations}

The dimensionless groups are given by:

\begin{equation}
\begin{alignedat}{1}Q & =\frac{e^{2}\mu^{2}}{L^{3}mG^{2}}\\
V & =\frac{eV_{0}\mu^{2}}{hLmG^{2}}\\
F_{ve} & =\frac{\pi r_{0}^{2}\mu^{2}}{mLG}
\end{alignedat}
\end{equation}
The three group measure the strength of Coulomb interactions, external
field and viscoleastic forces in units
of the inertial force $m d \upsilon/dt$, respectively.
In these units, $F_{ve}=Q$, so that we are left with four
independent parameters (considering also $\bar{\alpha}$ and $\bar{D}_{\upsilon}$).

\subsection{Numerical integration}

The equations of motion equations are discretised on a uniform sequence
$t_{i}=t_{0}+i\Delta t$, $i=1,\ldots,n_{steps}$.
At each time step, we first
integrate the stochastic Eq (\ref{eq:ODE}c) using
the explicit ``strong'' order scheme proposed by E. Platen
\cite{platen1987derivative,kloeden1992numerical},
whereof the order of ``strong'' convergence was evaluated in literature
equal to $1.5$. 
 
The Platen scheme delivers:

\begin{equation}
\begin{alignedat}{1}Y_{t+\varDelta t}^{k}= & Y_{t}^{k}+b^{k}\Delta\Omega+\frac{1}{2\left(\Delta t\right)^{\sfrac{1}{2}}}\left[a^{k}\left(\vec{\boldsymbol{\Upsilon}}_{+}\right)-a^{k}\left(\vec{\boldsymbol{\Upsilon}}_{-}\right)\right]\Delta\varPsi\\
 & +\frac{1}{4}\left[a^{k}\left(\vec{\boldsymbol{\Upsilon}}_{+}\right)+2a^{k}+a^{k}\left(\vec{\boldsymbol{\Upsilon}}_{-}\right)\right]\Delta t
\end{alignedat}
\end{equation}
with $Y_{t}^{k}$ denoting the approximation for the \textit{k}-th
component of a generic vector $\vec{\textbf{X}}$ whereof the time derivative
is $d\vec{\textbf{X}}/dt=\vec{\textbf{a}}\left(t,X^{1},\ldots,X^{d}\right)+\vec{\textbf{b}}\, d\Omega$, 
denoting $\Omega\left(t\right)$ a Wiener process. Here, the vector supporting
values $\vec{\boldsymbol{\Upsilon}}_{\pm}$ are 

\begin{equation}
\vec{\boldsymbol{\Upsilon}}_{\pm}=\vec{\textbf{Y}}_{t}+\vec{\textbf{a}}\Delta t\pm\vec{\textbf{b}}\left(\Delta t\right)^{\sfrac{1}{2}},
\end{equation}
and $\Delta\Omega$ and $\Delta\varPsi$ are normally distributed
random variables related to two independent $N\left(0,1\right)$ standard
Gaussian distributed random variables $U_{1}$ and $U_{2}$ via the
linear transformation:

\begin{equation}
\Delta\Omega=U_{1}\left(\Delta t\right)^{\sfrac{1}{2}}\qquad\Delta\varPsi=\frac{1}{2}\left(\Delta t\right)^{\sfrac{3}{2}}\left(U_{1}+\frac{1}{\sqrt{3}}U_{2}\right).\label{eq:random-var-1}
\end{equation}

Note that for $\left|\vec{\textbf{b}}\right|=0$ the Platen scheme reduces to
the second-order Runge-Kutta scheme. Finally, the remaining Eqs (\ref{eq:ODE}a)
and (\ref{eq:ODE}b) are integrated via a second order Runge-Kutta
scheme with the $\bar{W}_{i+1}$ value previously obtained via the
Platen scheme.

\section{Results and Discussion}

We investigate different regimes of electrified jets associated with
the Eqs (\ref{eq:SDE-dimen-new}), with special focus on metastable
states and asymptotic behavior. 
As a reference case, we consider the
typical values of $Q=12$ and $V=2$, already investigated in previous
works\cite{pontrelli2014electrospinning,reneker2000bending}. 
All simulations start from the same initial conditions: $\bar{l}=1$, $\bar{\sigma}=0$
and $\bar{W}=0$. Note that for reference parameters developed by
experimental results\cite{reneker2000bending} the typical values
of length scale $L$ and relaxation time $\tau$ are $3.19\:\text{\text{mm}}$,
and $10^{-2}\,\text{s}$, respectively. 

\subsection{Parameter setup and asymptotes}
First, we study the deterministic case, by imposing 
$\bar{\alpha}=\bar{D}_{\upsilon}=0$. 
We integrate in time forward and backward Eqs (\ref{eq:ODE}) in the
interval $\bar{t}_{a}=0$ and $\bar{t}_{b}=5$ at different values
of time-step $\Delta\bar{t}$ in order to assess a suitable value for the specific case under investigation. 
Exploiting the time reversibility, we measured an average
absolute error $\Delta\bar{l}=\left|\bar{l}_{2n_{steps}}-\bar{l}_{0}\right|$
lower than $10^{-12}$ with time step $\Delta\bar{t}=10^{-2}$.
Therefore, we take a time step $\Delta\bar{t}=10^{-3}$, as a conservative
choice for the specific case under investigation.

We next discuss the elongation of the jet under stochastic perturbation.
To estimate the parameters $\bar{\alpha}$, $S$ and $P$, we consider
the empirical formula Eq (\ref{eq:emp-air}) for the dissipative air
drag force\cite{spinning1991science,sinha2010meltblowing,yarin2014fundamentals}.
For typical air density, kinematic viscosity of air gaseous medium
and jet mass (assumed constant), we obtain a $\bar{\alpha} \sim 0.5$ 
(with $\tau=10^{-2}\,\text{s}$), while the parameters
$S$ and $P$ are set equal to $0.905$ and $0.19$, respectively.
For all the investigated cases, we adopt for the sake of simplicity the value $\bar{D}_{\upsilon}=\bar{\alpha}$.
In order to accumulate sufficient statistics, we have run $10000$ independent
trajectories for each different case under investigation. Thence,
we compute the time dependent mean value of our observables along
the dynamics.

We integrate the Eqs (\ref{eq:SDE}) for three different cases: in the
first, we set $\bar{\alpha}=0$, $S=0$ and $P=0$ (\textit{deterministic
case}); in the second $\bar{\alpha}=0.5$, $S=0$ and $P=0$ (\textit{linear
Langevin}); for the third case (\textit{non-linear Langevin}) $\bar{\alpha}=0.5$
while and $S=0.905$ and $P=0.19$, respectively.

\begin{figure}[H]
\begin{centering}
\includegraphics[scale=0.35]{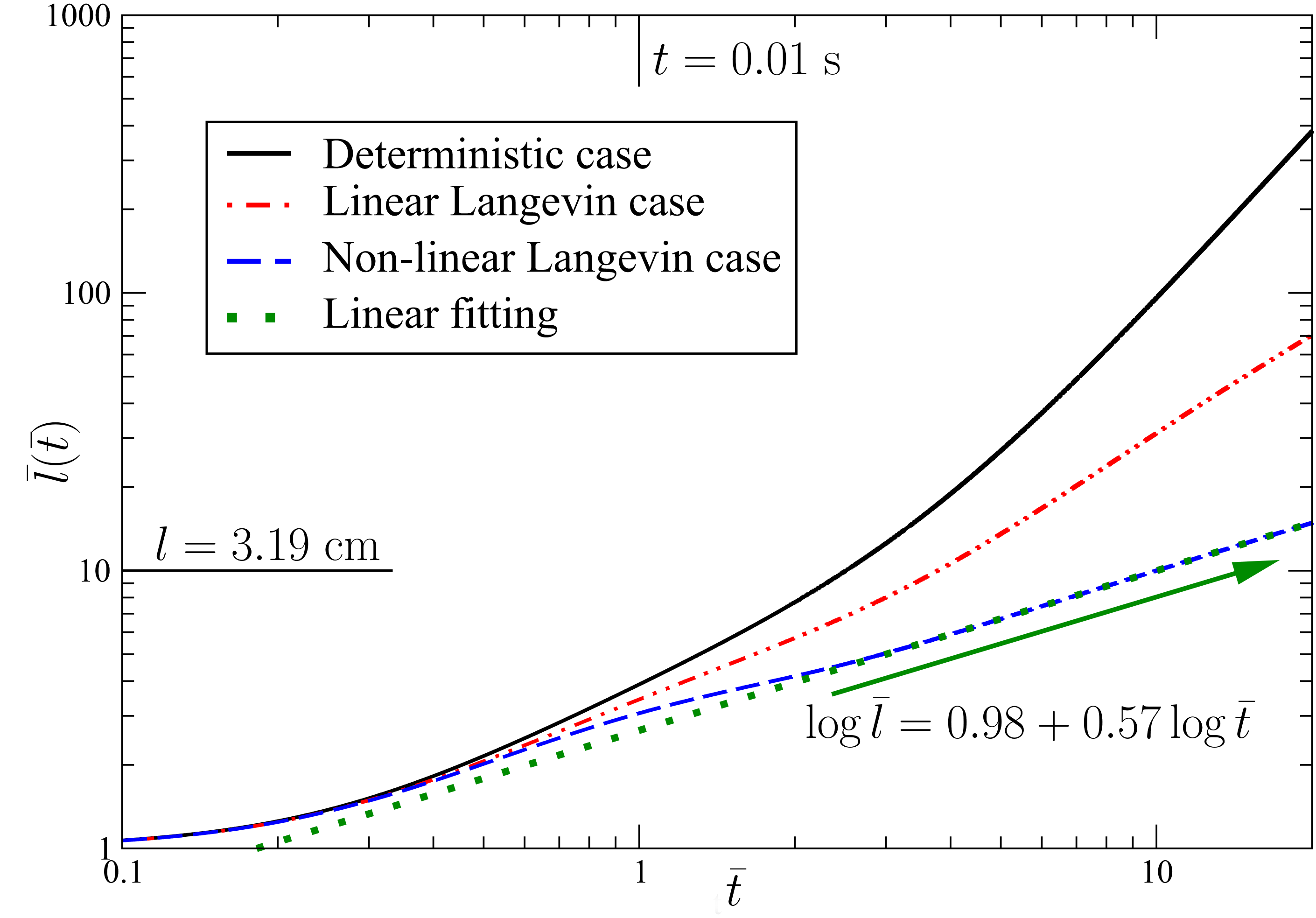} 
\par\end{centering}

\protect\caption{{\small{}Time evolution of the jet elongation $\bar{l}\left(\bar{t}\right)$
for three different cases: 1) Deterministic: $\bar{\alpha}=0$, $S=0$
and $P=0$ in black continuous line; 2) Linear Langevin: $\bar{\alpha}=0.5$,
$S=0$ and $P=0$ in red dashed line; 3) Non-linear Langevin: $\bar{\alpha}=0.5$,
$S=0.905$ and $P=0.19$ in blue dotted-dashed line. In green dotted line we report the linear fitting (in log-log scale) 
for the Non-linear Langevin case with slope equal to the expected value $\sfrac{4}{7}$. The horizontal line
on the left side corresponds to the physical length $\l=3.19$
cm. }}

\label{Fig:case-confr-l} 
\end{figure}

\begin{figure}[H]
\begin{centering}
\includegraphics[scale=0.35]{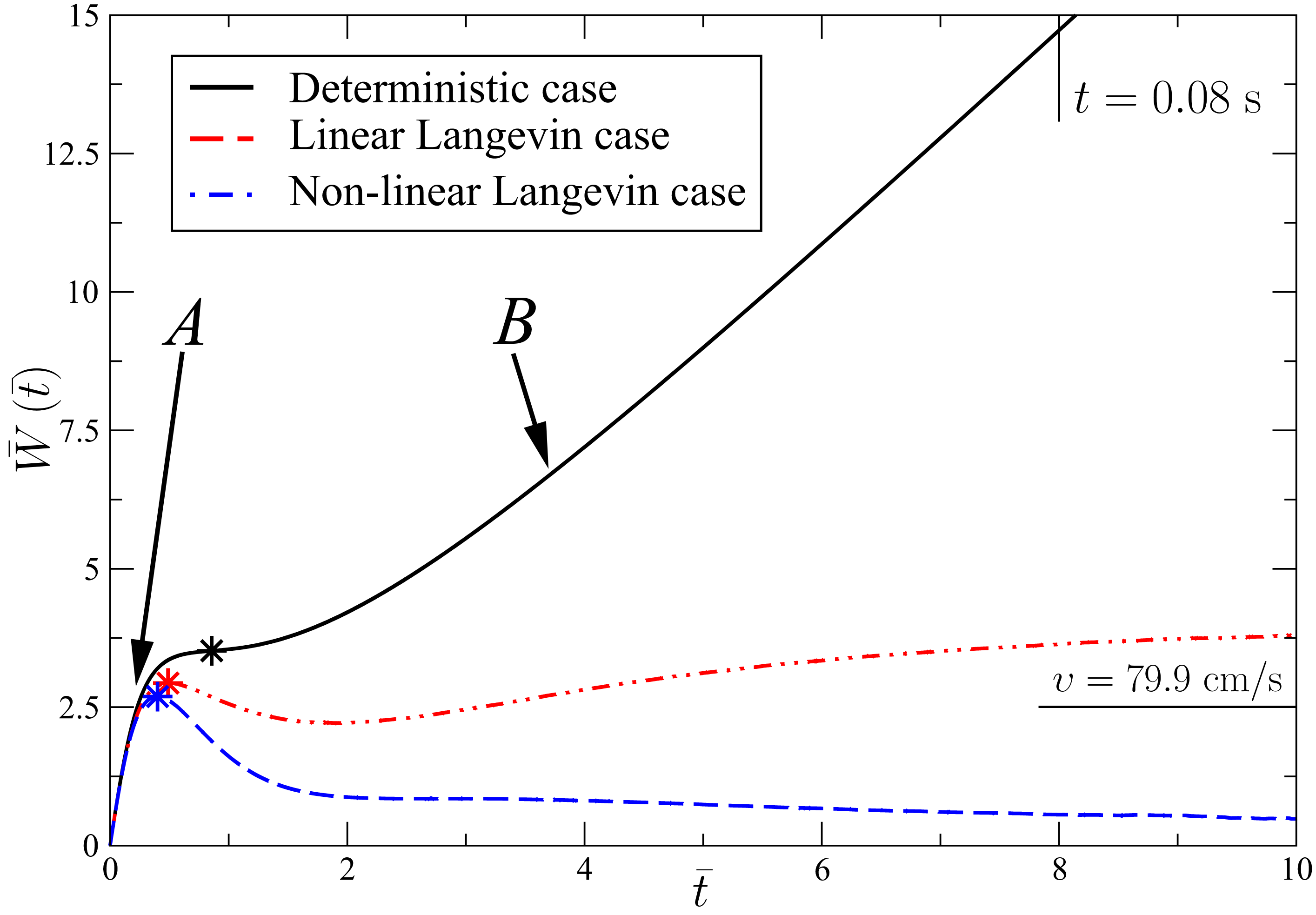} 
\par\end{centering}

\protect\caption{{\small{}Time evolution of the velocity $\bar{W}\left(\bar{t}\right)$
for three different cases: 1) Deterministic: $\bar{\alpha}=0$, $S=0$
and $P=0$ in black continuous line; 2) Linear Langevin: $\bar{\alpha}=0.5$,
$S=0$ and $P=0$ in red dashed line; 3) Non-linear Langevin: $\bar{\alpha}=0.5$,
$S=0.905$ and $P=0.19$ in blue dotted-dashed line. Two stages of
the elongation process are identified: an early transient ($A$), which
comes to a quasi stationary point (denoted by a star symbol) and a
later stage ($B$) controlled by the competition between the external
field and air dissipation. The figure highlights the major role played
by dissipation on the long-term evolution of the system, which is
from free-fall acceleration to a constant velocity regime. As expected,
the super-linear dissipation enhances the drag effect, leading to a
further reduction of the time-asymptotic speed. The horizontal line
on the right side corresponds to the physical speed $\upsilon=79.9$
cm/s. }}

\label{Fig:case-confr} 
\end{figure}

Three basic time-asymptotic regimes can be identified:

i) Deterministic:  $\bar{W}\propto\bar{t}^{2}$, $\bar{l}\propto\bar{t}^{2}$.
This is the free-fall regime driven by the external voltage, once
every other force is extinguished (see Fig \ref{Fig:case-confr-l}).

ii) Stochastic, Linear Dissipation: $\bar{W} \propto const$, $\bar{l} \propto \bar{t}$. 
This is the ballistic regime resulting from the balance between
the external field and linear dissipation.
Asymptotically, the jet moves at a constant speed, like electrons
in a linear host media.

iii) Stochastic, Non-linear: $\bar{l} \propto \bar{t}^{4/7}$, $\bar{W} \propto \bar{t}^{-3/7}$.
This is the regime resulting from the balance between
the external field and non-linear dissipation with $S=0.905$ and $P=0.19$.
Note that, even though these exponents are close to the case of standard
diffusion, they result from a very different process, namely a constant
force against a drag growing with the filament length.
It is interesting to notice that the filament length is still unbounded
in the limit $t \to \infty$ (see Fig \ref{Fig:case-confr-l}), 
but much slower than the ballistic case
associated with linear dissipation.
We emphasize that these asymptotic regimes, although
important for a qualitative analysis of the process, bear limited
practical interest since, in the long-term, the jet undergoes a 
bending instability which cannot be described by the present 
one-dimensional model. The one-dimensional model is however very useful
to discuss the early-stage of the evolution and incipient onset of the
instability.
In Fig \ref{Fig:case-confr}, we report the velocity $\bar{W}\left(\bar{t}\right)$
versus time for the three cases above. 
\subsection{Deterministic case: no dissipation}
In the deterministic case,
we identify two sequential stages in the elongation process (denoted
$A$ and $B$ in Fig \ref{Fig:case-confr}). In the first regime, we observe a small
increase of $\bar{W}\left(\bar{t}\right)$, which rises up to achieve
a quasi stationary point denoted by $\bar{t}_{*}$, where the viscoelastic
force ${\displaystyle \frac{F_{ve}\bar{\sigma}\left(\bar{t}_{*}\right)}{\bar{l}\left(\bar{t}_{*}\right)}}$
balances the sum of the two force terms 
${\displaystyle \frac{Q}{\bar{l}\left(\bar{t}_{*}\right)^{2}}}$
and $V$, providing a zero total force. 
Subsequently, after about $20-40$ ms, the velocity attains a near-linearly
increasing trend, close to the time-asymptotic solution discussed above 
(see also Ref \cite{pontrelli2014electrospinning}). 
Note that the instant $\bar{t}_{*}$ corresponds
to the lower limit of the derivative 
$\partial\bar{W}\left(\bar{t}\right)/\partial\bar{t}$,
and discerns the two stages of the dynamics, regime $A$ characterized
by the competition between Coulomb and viscoelasticity, and regime $B$,
characterized by the sole action of the external field.
\subsection{Stochastic case: linear dissipation}
Once linear dissipation is included, the jet elongation suffers
an additional slow down, leading to a decrease of the velocity
after the early peak driven by Coulomb forces.
This leads to a local minimum at $\bar{t}_{**}$  with no counterpart
in the deterministic case, as already discussed in Ref \cite{lauricella2014electrospinning}. 
We observe that the time occurrence $\bar{t}_{*}$ of the peak
is anticipated by the air drag, an the corresponding velocity slightly decreased. 
Subsequently, in the regime $B$ the velocity attains its
asymptotic value (see Fig \ref{Fig:case-confr}).
\subsection{Stochastic case: non-linear dissipation}
In the third case (non-linear Langevin), we observe that the non-linearity
of the dissipative term largely alters the time evolution of velocity. 
In particular, it leads to just one quasi stationary point. Furthermore,
the velocity $\bar{W}\left(\bar{t}\right)$ appears to tend asymptotically
to zero, as it pertains to its asymptotic regime

\begin{figure}
\begin{centering}
\includegraphics[scale=0.35]{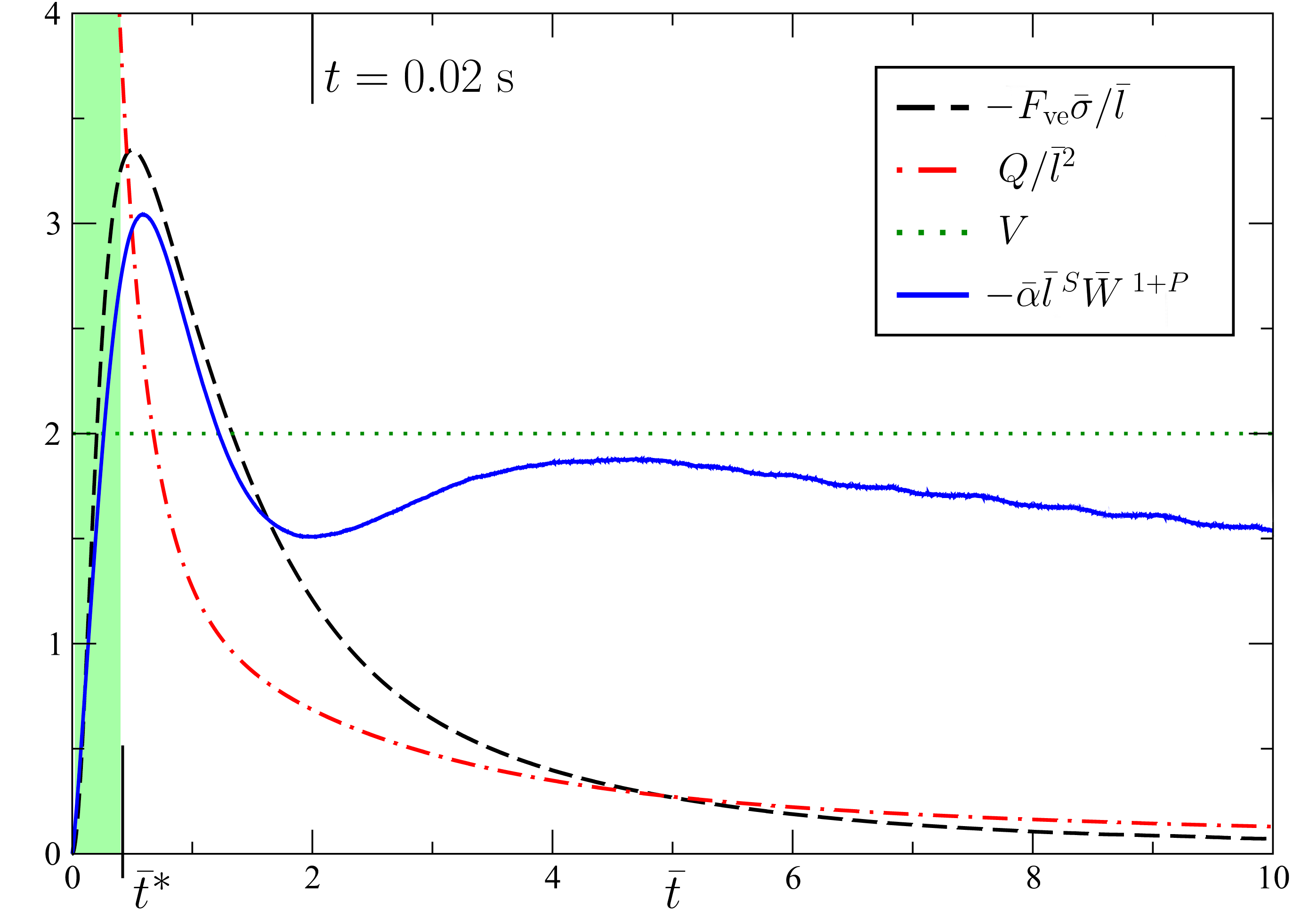} 
\par\end{centering}

\protect\caption{{\small{}Time evolution of the four force terms versus time $\bar{t}$:
$-{\displaystyle\bar{\alpha}\bar{l}^{\, S}\bar{W}^{1+P}}$ (continuous line), $-{\displaystyle \frac{F_{ve}\bar{\sigma}}{\bar{l}}}$
(dashed line), $V$ (dotted line), and ${\displaystyle \frac{Q}{\bar{l}^{2}}}$
(dashed-dotted line) for the non-linear Langevin case $\bar{\alpha}=0.5$.
The quasi-stationary point $\bar{t}^{*}=0.4$ is highlighted.
The viscoelastic force $-{\displaystyle \frac{F_{ve}\bar{\sigma}}{\bar{l}}}$
peaks at about $\bar{t}=0.5$. Subsequently, the force terms ${\displaystyle -\frac{F_{ve}\bar{\sigma}}{\bar{l}}}$
and ${\displaystyle \frac{Q}{\bar{l}^{2}}}$ decay to zero, while
the term $-\bar{\alpha}\bar{l}^{S}\bar{W}^{1+P}$ and the external electrical
field $V$ govern the jet dynamics. }}

\label{Fig:stoc-force} 
\end{figure}

We next examine the force terms, as
shown in Fig \ref{Fig:stoc-force}. 
First, the Coulombic term ${\displaystyle \frac{Q}{\bar{l}\left(\bar{t}\right)^{2}}}$
decays rapidly so that it plays no role in regime $B$. 
The early-stage of the dynamics is characterized by the terms ${\displaystyle \frac{F_{ve}\bar{\sigma}\left(\bar{t}\right)}{\bar{l}\left(\bar{t}\right)}}$
and $\bar{\alpha}\bar{l}^{\, S}\bar{W}^{1+P}$, which increase as
consequence of the jet stretching due to the external electric field.
This early-stage comes to the quasi stationary point at time $\bar{t}_{*}$,
where  
${\displaystyle \frac{F_{ve}\bar{\sigma}\left(\bar{t}\right)}{\bar{l}\left(\bar{t}\right)}}$
and $\bar{\alpha}\bar{l}^{\, S}\bar{W}^{1+P}$ conspire to balance 
the external drive $V$.

After the instant $\bar{t}_{*}$, the term $\bar{\alpha}\bar{l}^{\, S}\bar{W}^{1+P}$
becomes larger in magnitude, leading to a decrease of the
velocity $\bar{W}$, due to the dissipative and viscoelastic terms
(see Fig \ref{Fig:case-confr}). 
At the same time, the term ${\displaystyle \frac{F_{ve}\bar{\sigma}\left(\bar{t}\right)}{\bar{l}\left(\bar{t}\right)}}$
starts to decay, and the jet dynamics is governed only by the remaining
opposite terms $V$ and $\bar{\alpha}\bar{l}^{\, S}\bar{W}^{1+P}$.
Since dissipation grows quasi-linearly with the jet elongation, the
velocity goes to zero in the time asymptotic limit. 

\begin{figure}
\begin{centering}
\includegraphics[scale=0.35]{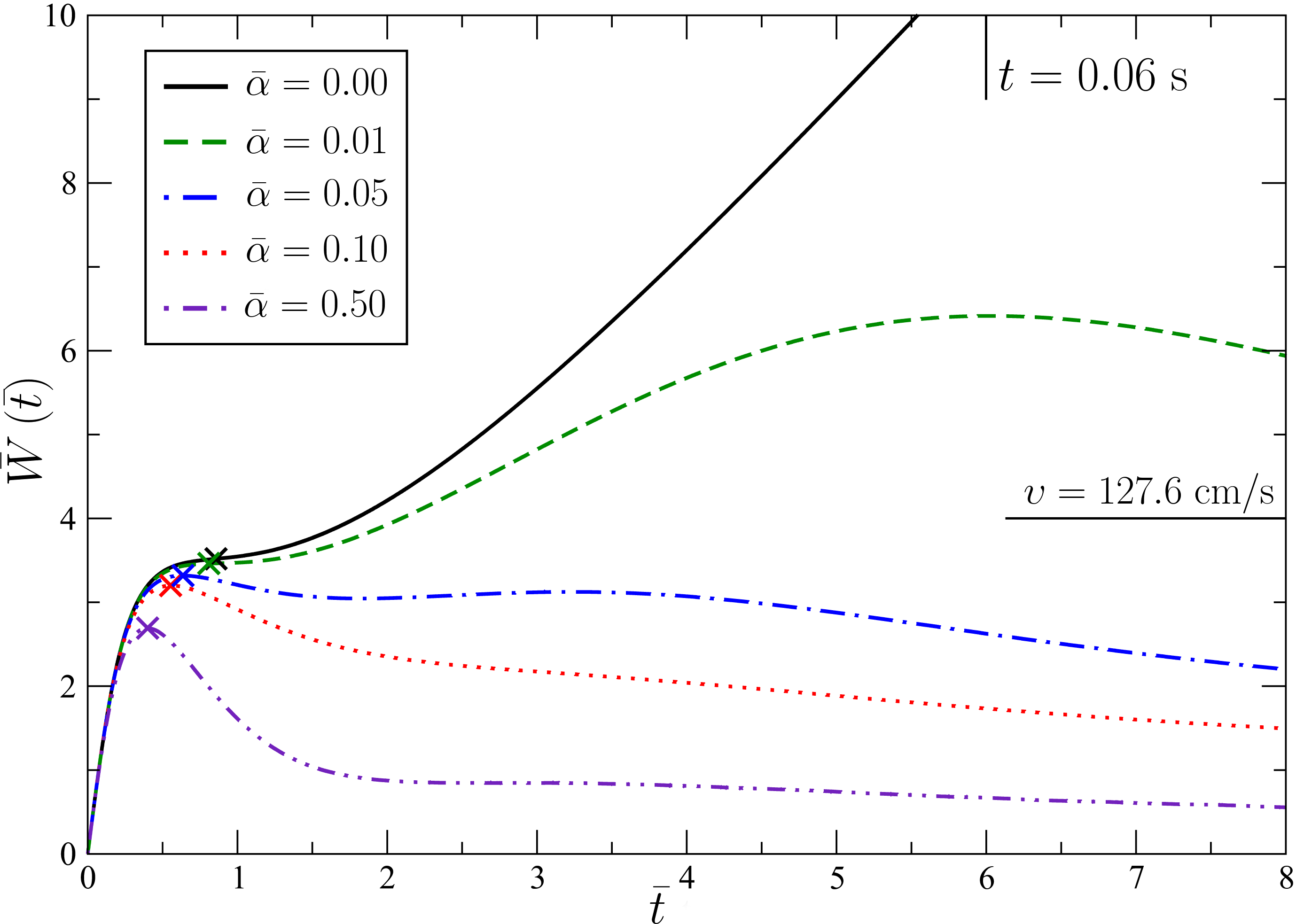} 
\par\end{centering}

\protect\caption{{\small{}Time evolution of the velocity $\bar{W}\left(\bar{t}\right)$
for different values of $\bar{\alpha}$. From top to bottom curves:
$\bar{\alpha}$ = 0 (continuous line), 0.01 (dashed line), 0.05 (dashed-dotted
line), and 0.1 (dotted line), and 0.5 (dashed-dotted-dotted line),
keeping $\bar{D}_{\upsilon}=\bar{\alpha}$ for all the cases. The
quasi-stationary points are denoted by a star symbol. As expected,
increasing the drag coefficient entails a substantial reduction of
the filament velocity. The horizontal line on the right side corresponds
to the physical speed $\upsilon=127.6$ cm/s. }}

\label{fig:vel-stoc} 
\end{figure}

Next, we investigate the elongation of jet under stochastic perturbation
modelled by the non-linear Langevin equation for different values
of $\bar{\alpha}$, keeping the condition $\bar{D}_{\upsilon}=\bar{\alpha}$.
In particular, we explore the way that the position of $\bar{t}_{*}$
is altered by the dissipative-perturbing force $-\bar{\alpha}\bar{l}^{\, S}\bar{W}^{1+P}+\sqrt{2\bar{D}_{\upsilon}}\eta\left(\bar{t}\right)$
in Eq (\ref{eq:SDE}). In Fig \ref{fig:vel-stoc}, we report the time
evolution of the velocity $\bar{W}\left(\bar{t}\right)$ for different
$\bar{\alpha}$. For all investigated regimes we observe a quasi
stationary point at time $\bar{t}_{*}$, which decreases by increasing the
term $\bar{\alpha}$ (see Tab \ref{tab:PQSvalues}).

\begin{figure}
\begin{centering}
\includegraphics[scale=0.35]{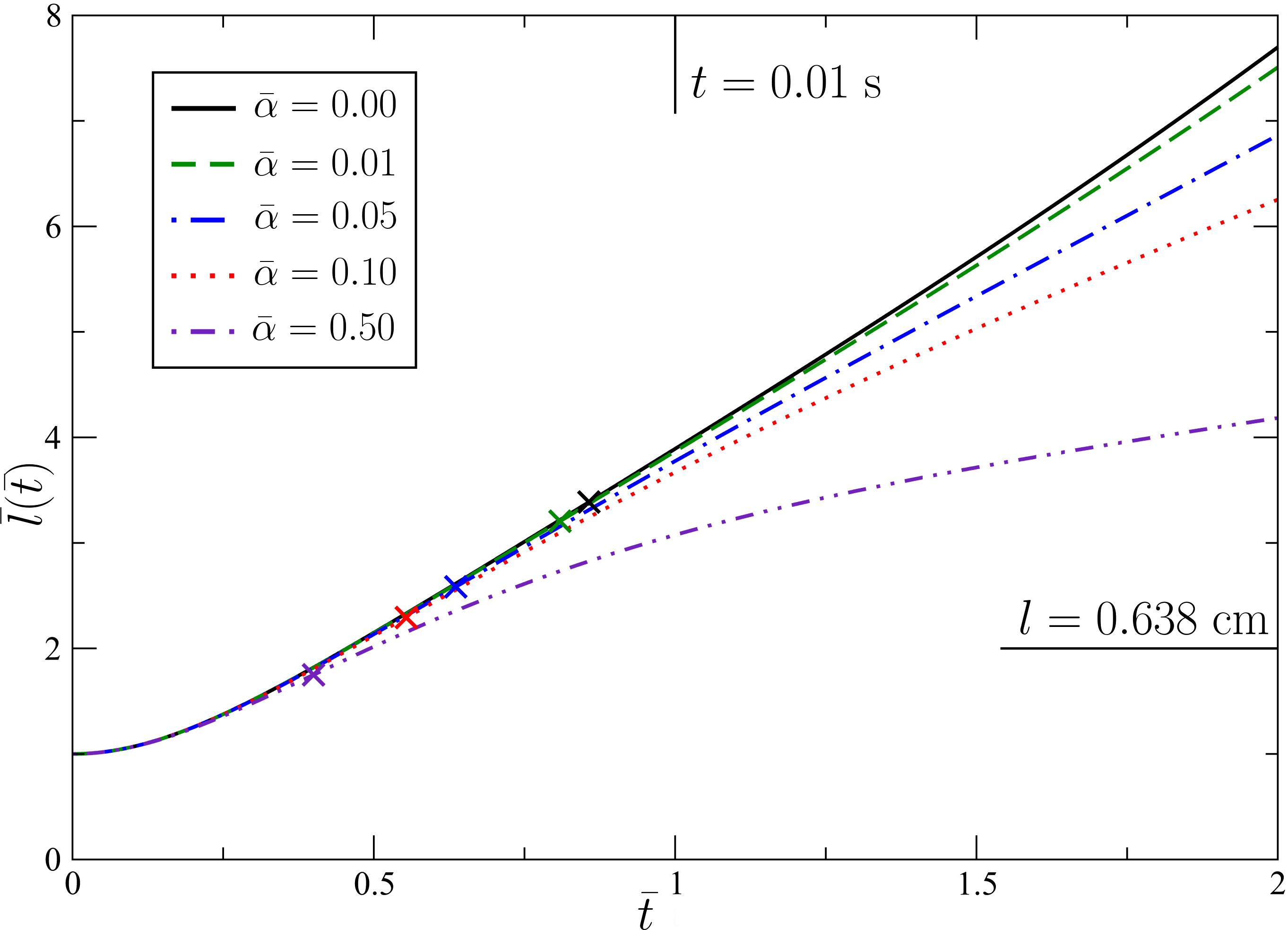} 
\par\end{centering}

\protect\caption{{\small{}Time evolution of the jet elongation $\bar{l}\left(\bar{t}\right)$
for different values of $\bar{\alpha}$: $0$ (continuous line), $0.01$
(dashed line), $0.05$ (dashed-dotted line), and $0.1$ (dotted line),
and $0.5$ (dashed-dotted-dotted line), keeping $\bar{D}_{\upsilon}=\bar{\alpha}$
for all the cases. The quasi-stationary points are depicted as star
symbol. The horizontal line on the right side corresponds to the
physical length $l=0.638$ cm.}}

\label{Fig:caso-stoc-l} 
\end{figure}

The straight path of the electrified jet is described by the observable
$\bar{l}\left(\bar{t}_{*}\right)$, which is seen to decrease by increasing
$\bar{\alpha}$ (Fig \ref{Fig:caso-stoc-l}). 
Useful hints for the optimal design of the electrospinning processes,
resulting from deeper insights into the early-stage dynamics of the
jet can be numerous, including the possibility of better controlling
the subsequent development of three-dimensional instabilities, and
consequently, the diameter and morphology of collected nanostructures,
as well as the assembly and positions of nanofibers impinging onto
the collector. A better control of these nanofabrication processes
could therefore entail the identification and tailoring of air drag
mechanisms, which can eventually be induced by a proper designed of
gas-injecting systems nearby the spinneret.

\section{Conclusions}

Summarizing, we have investigated the flow of charged viscoelastic
fluids in the presence of stationary stochastic perturbations. A Brownian
term has been used to model the effects of a perturbing force on the
stretching properties of electrically charged jets, providing significant
qualitative new insights. We demonstrated that the non-linear
dependence of the dissipative term on the geometry of the electrospun
polymer provides important effects that cannot be properly modelled
by a linear Langevin-like stochastic differential equation. 
We also observed that the air drag force significantly affects the dynamics
of the electrospinning process, leading to a time-asymptotic vanishing velocity of the jet. 
Furthermore, a reduction of the linear extension
of the jet is observed in the early-stage by increasing the dissipative
force term. These results may contribute to the optimal set-up
of the experimental conditions, so as to enhance the efficiency
of the process and the quality of the electrospun fibers. 
These may include, among others, environmental vibrations and 
resulting micro-vorticity patterns.

\section*{Acknowledgments}

This work is dedicated to Jean-Pierre Hansen, on the occasion of his
70th birthday. One of the authors (SS) would like to thank Jean-Pierre for
many stimulating discussions on many topics in 
statistical mechanics at large.
Whether in Rome, Cambridge or anywhere else, conversations with Jean-Pierre
have always been highly enriching on both scientific and human sides. 
The research leading to these results has received funding from the European Research 
Council under the European Union's Seventh Framework Programme (FP/2007-2013)/ERC 
Grant Agreement n. 306357 ("NANO-JETS").

\newpage{}

\section*{Tables}

\begin{table}[H]
\begin{centering}
\begin{tabular}{ccccc}
\hline 
$\bar{\alpha}$  & $\bar{t}_{*}$  & $\bar{l}\left(\bar{t}_{*}\right)$  & $\bar{\sigma}\left(\bar{t}_{*}\right)$  & $\bar{W}\left(\bar{t}_{*}\right)$ \tabularnewline
\hline 
\hline 
0  & 0.86  & 3.39  & 0.81  & 3.52 \tabularnewline
0.01  & 0.81  & 3.21  & 0.80  & 3.46 \tabularnewline
0.05  & 0.64  & 2.58  & 0.71  & 3.32 \tabularnewline
0.1  & 0.55  & 2.29  & 0.65  & 3.20\tabularnewline
0.5  & 0.40  & 1.75  & 0.47  & 2.69 \tabularnewline
\hline 
\end{tabular}
\par\end{centering}

\caption{Values of the dimensionless variables length $\bar{l}$,
stress $\bar{\sigma}$ and velocity $\bar{W}$ at the quasi-stationary
points $\bar{t}_{*}$ computed for different values of $\bar{\alpha}$. 
To be noted the decrease of $\bar{t}_{*}$ at increasing
$\bar{\alpha}$ .Furthermore, we stress the decrease of the lenght
$\bar{l}\left(\bar{t}_{*}\right)$ (see Fig. \ref{Fig:caso-stoc-l}),
proving that the initial stage of the elongation process contracts
as a consequence of the uniaxial perturbation.}

\label{tab:PQSvalues} 
\end{table}

\section*{Figures}

\newpage{}

\section*{Supplemental Information}

\setcounter{equation}{0}

In this paper, we use a recast form of the empirical formula for the
air drag force $f_{air}$ usually found in literature%
\footnote{A. Ziabicki and H. Kawai, \textquotedbl{}High-Speed Fiber Spinning:
Science and Engineering Aspects\textquotedbl{}, Krieger Publishing
Co (1991); Sinha-Ray, Suman and Yarin, Alexander L and Pourdeyhimi,
Behnam, \textquotedbl{}Meltblowing: I-basic physical mechanisms and
threadline model\textquotedbl{}, Journal of Applied Physics (2010),
034912; Yarin, Alexander L and Pourdeyhimi, Behnam and Ramakrishna,
Seeram, \textquotedbl{}Fundamentals and Applications of Micro and
Nanofibers\textquotedbl{}, Cambridge University Press (2014).%
}. Here, we provide few details on our revised expression. The original
empirical formula for the air drag force $f_{air}$ is

\begin{equation}
f_{air}=l\cdot0.65\pi r\rho_{air}\upsilon^{2}\left(\frac{2\upsilon r}{\nu_{air}}\right)^{-0.81},
\end{equation}
where we consider $l$ as the distance between the beads labeled $i$
and $i-1$ , $r$ the cross-sectional radius of the filament, $\upsilon$
is the velocity of the $i-th$ bead, $\rho_{a}$ is the air density,
and $\nu_{a}$ is the air kinematic viscosity.

We rearrange the Eq as

\begin{equation}
f_{air}=l\cdot0.65\pi\rho_{air}\left(\frac{2}{\nu_{air}}\right)^{-0.81}r^{0.19}\upsilon^{1.19}.
\end{equation}

Assuming a constant volume of the jet $\pi r^{2}l=\pi r_{0}^{2}L$,
so that $r=r_{0}\sqrt{L/l}$ with $L$ and $r_{0}$ respectively the
length and the radius of the jet segment between the beads $i$ and
$i-1$ at the nozzle before the stretching, we obtain

\begin{equation}
f_{air}=0.65\pi\rho_{air}\left(\frac{2}{\nu_{air}}\right)^{-0.81}l\left(r_{0}L^{0.5}l^{-0.5}\right)^{0.19}\upsilon^{1.19}.
\end{equation}

Thus, we can write

\begin{equation}
f_{air}=\left[0.65\pi\rho_{air}\left(\frac{2}{\nu_{air}}\right)^{-0.81}L^{0.095}r_{0}^{0.19}\right]l^{0.905}\upsilon^{1.19}.
\end{equation}

If we write $f_{air}$ as

\begin{equation}
f_{air}=m_{i}\alpha l^{0.905}\upsilon^{1+0.19},
\end{equation}

denoting $m_{i}$ the mass of the bead $i$, we obtain an equivalent
formula for the air drag force with the dissipative coefficient $\alpha$
equal to

\begin{equation}
\alpha=0.65\pi\rho_{air}\left(\frac{2}{\nu_{air}}\right)^{-0.81}\frac{L^{0.095}r_{0}^{0.19}}{m_{i}}.
\end{equation}

\end{document}